\documentclass[12pt, onecolumn, notitlepage]{article}
\usepackage{hyperref}
\usepackage{graphicx}
\usepackage{pdfpages}
\usepackage{times}
\usepackage{xcolor}
\usepackage{amssymb}
\selectfont

\begin{document}
\title{\vspace{-65pt}Epidemic growth and Griffiths effects on an emergent network of excited atoms}


\author{
T. M. Wintermantel$^{1,2}$, M. Buchhold$^3$, S. Shevate$^2$, M. Morgado$^2$,\\Y. Wang$^2$, G. Lochead$^2$, S. Diehl$^3$, S. Whitlock$^{2\ast}$ \\
\\
\normalsize{$^{1}$Physikalisches Institut, Universit\"at Heidelberg, 69120 Heidelberg, Germany}\\
\normalsize{$^{2}$ISIS (UMR 7006), University of Strasbourg and CNRS, 67000 Strasbourg, France}\\
\normalsize{$^{3}$Institut f\"ur Theoretische Physik, Universit\"at zu K\"oln, 50923 Cologne, Germany}\\
\\
\normalsize{$^\ast$Corresponding author. E-mail:  whitlock@unistra.fr.}
}

\date{\today}

\maketitle
\textbf{
Whether it be physical, biological or social processes, complex systems exhibit dynamics that are exceedingly difficult to understand or predict from underlying principles~\cite{barrat2008dynamical}. Here we report a striking correspondence between the collective excitation dynamics of a laser driven ultracold gas of Rydberg atoms and the spreading of diseases, which in turn opens up a highly controllable experimental platform for studying non-equilibrium dynamics on complex networks~\cite{PastorSatorras2015}. We find that the competition between facilitated excitation and spontaneous decay results in a fast growth of the number of excitations that follows a characteristic sub-exponential time dependence which is empirically observed as a key feature of real epidemics\cite{Viboud2016,Chowell2016mathematical}. Based on this we develop a quantitative microscopic susceptible-infected-susceptible (SIS) model which links the growth and final excitation density to the dynamics of an emergent heterogeneous network and rare active region effects associated to an extended Griffiths phase~\cite{Moreira1996,Voltja2005,Munoz2010}. This provides physical insights into the nature of non-equilibrium criticality in driven many-body systems and the mechanisms leading to non-universal power-laws in the dynamics of complex systems.}
 

The dynamical behavior of an exceptionally diverse spectrum of real-world systems is governed by critical events and phenomena occurring on vastly different spatial and temporal scales. A disease outbreak, for example, can be very sensitive to the type of disease and the behavior of individuals, yet epidemics generically feature a characteristic time dependence~\cite{Viboud2016} that emerges from the connections within and between communities~\cite{PastorSatorras2015,Chowell2016mathematical}. In studying these systems, complex networks provide a crucial layer of abstraction to bridge the behavior of individuals and the macroscopic consequences. Accordingly, they have found applications not only in biology and the study of epidemics~\cite{PastorSatorras2015}, but also in informatics~\cite{Kephart1992directed}, marketing~\cite{Bampo2008effects}, finance~\cite{Peckham2014contagion}, and traffic flow~\cite{saberi2020simple}. An overarching challenge in these fields is to find general principles governing complex system dynamics and to pinpoint how apparent universal characteristics emerge from the underlying network structure.

\begin{figure}[!ht]%
    \centering
	\includegraphics[width=6.25in]{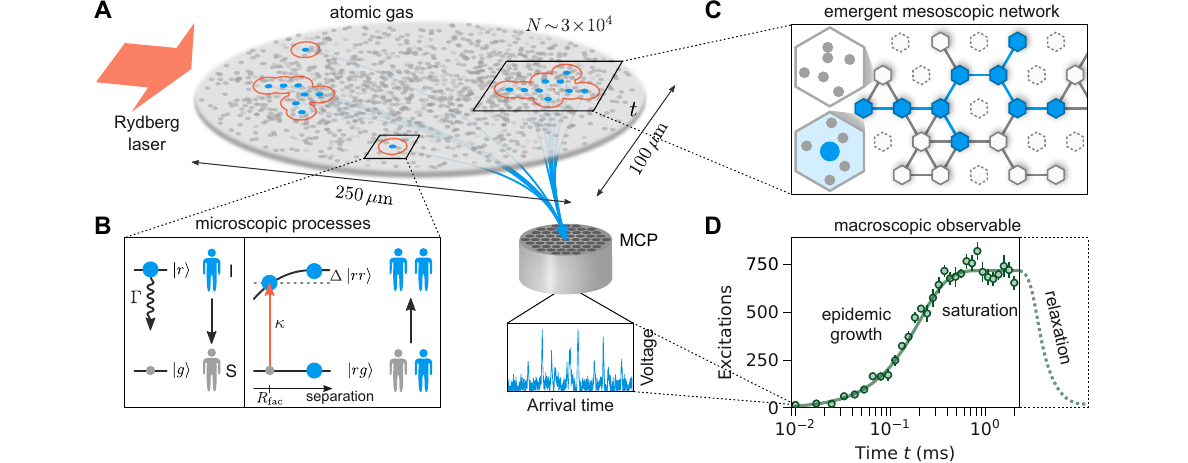}
	\caption{\footnotesize 
	\textbf{Physical system for studying epidemic growth and dynamics on complex networks.} 
	\textbf{A} Experiments are performed on a two-dimensional gas of potassium atoms driven by an off-resonant laser field. The gas is initially prepared with a small number of seed excitations (blue disks) which then evolves according to the microscopic processes depicted in B, giving rise to growing excitation clusters that spread throughout the system. After different exposure times $t$, the Rydberg atoms are field ionized and detected on a microchannel plate detector (MCP), where the incident ions create voltage spikes (blue trace).
	\textbf{B} Each atom can be treated as a two-level system with a ground state $|g\rangle$ (gray disks) and excited Rydberg state $|r\rangle$ (larger blue disks). Excited atoms can decay with rate $\Gamma$ or facilitate additional excitations with rate $\kappa$ at a characteristic distance $R_\mathrm{fac}$ analogous to the transmission of an infection. 
	\textbf{C} The dynamics of this system can be described by a susceptible-infected-susceptible (SIS) model on an emergent heterogeneous network. Each node represents a discrete cell of the coarse grained system, which can be infected (with excitation, blue) or susceptible (without excitation, white). The infection probability of each node is weighted by the number of atoms in that cell that can undergo facilitated excitation (disconnected nodes corresponding to vacant cells are depicted with dashed lines).
	\textbf{D} Exemplary data and numerical simulations (solid line) showing three different stages of dynamics: rapid growth, saturation, and eventual relaxation (data not shown). Error bars represent the standard error of the mean over typically $16$ experiment repetitions.
	}
	\label{fig:atomsnetwork}
\end{figure}

We address this challenge using a highly-controllable complex system that consists of a trapped ultracold atomic gas continuously driven to strongly interacting Rydberg states by an off-resonant laser field (Fig.~\ref{fig:atomsnetwork}). Our main findings include: (i) the rapid growth of excitations driven by a competition between microscopic facilitated excitation and decay processes (playing the role of the transmission of an infection and recovery respectively). The observed dynamics follow a power-law time dependence that parallels that which is empirically observed in real-world epidemics, providing a powerful demonstration of universality reaching beyond physics; (ii) a full description and interpretation of the experiment in terms of an emergent susceptible-infected-susceptible network linking the observed macroscopic dynamics to the microscopic physics; and (iii) the unexpected presence of rare region effects and a dynamical Griffiths phase associated to the emergent network structure, which gives rise to critical dynamics over an extended parameter regime and explains the appearance of power-law growth and relaxation, but with non-universal exponents.

\subsection*{Microscopic ingredients for an epidemic}
The microscopic processes governing the dynamics of ultracold atoms driven to Rydberg states by an off-resonant laser field, shown in Fig.~\ref{fig:atomsnetwork}(A,B), bear close similarities to those in epidemics~\cite{Perez-Espigares2017}. Each atom can be considered as a two-level system consisting of the atomic ground state (gray disks, healthy) and an excited Rydberg state (blue disks, infected). An excited atom can spontaneously decay (recovery, with rate $\Gamma$), or it can facilitate the excitation of other atoms (transmission of the infection, with rate $\kappa$) that satisfy certain constraints linked to their positions and velocities. This results in rapid spreading of the excitations through the gas (depicted by growing excitation clusters in Fig.~\ref{fig:atomsnetwork}A)~\cite{Ates2007,Schempp2014,Malossi2014,Urvoy2015,simonelli2016seeded}. 

Our experimental studies start from an ultracold thermal gas of $3\times 10^4$ potassium-39 atoms in their ground state $|g\rangle=4s_{1/2}$ which are held in a two-dimensional optical trap with a peak atomic density $n_{2D}(x\!=\!y\!=\!0)=0.76\,\mu$m$^{-2}$ (Fig.~\ref{fig:atomsnetwork}A). To trigger the dynamics at $t=0$ we apply a low intensity laser pulse tuned in resonance with the $|g\rangle \rightarrow |r\rangle=66p_{1/2}$ transition for a duration of $4\,\mu$s. This produces around 8 seed excitations at random positions within the gas. The laser is then suddenly detuned from resonance by $\Delta\nobreak=-30\,$MHz and adjusted in intensity. This makes it possible to facilitate secondary excitations at a distance $R_\mathrm{fac}=3.5\,\mu$m (illustrated by red circles), corresponding to the distance where the calculated Rydberg pair-state energy compensates the laser detuning. The two-body facilitation rate $\kappa$ is proportional to the laser intensity which can be tuned over a wide range. For the following measurements we choose different values of $\kappa$ ranging from $3.3\,$kHz to $10\,$kHz [see the Supplemental Material for details on the calibration of $\kappa$]. Additionally, Rydberg excitations spontaneously decay with a calculated rate $\Gamma=0.84\,$kHz (including black-body decay). Spontaneous (off-resonant) excitation events are very rare, with an estimated rate $\lesssim 1$\,kHz integrated over the whole cloud. To observe the system we measure the total number of Rydberg excitations present in the gas using field ionization and a microchannel plate detector for different exposure times $t$ up to $2\,$ms. Although each atom is identical and evolves according to these seemingly simple excitation rules, the competition between facilitated excitation and decay gives rise to complex dynamical phases and evolution~\cite{Lee2011,Lesanovsky2013,Carr2013,Gutierrez2017absorbing,Helmrich2018,HelmrichSOC}. However the full many-body system is even more complex: $3\times 10^4$ multilevel atoms moving in space with random positions and velocities while interacting with the laser field and each other; which makes it challenging to connect the microscopic physics to the macroscopic excitation dynamics~\cite{Helmrich2018,Goldschmidt2016,Marcuzzi2017}.

\begin{figure*}[!ht]%
    \centering
	\includegraphics[width=6.25in]{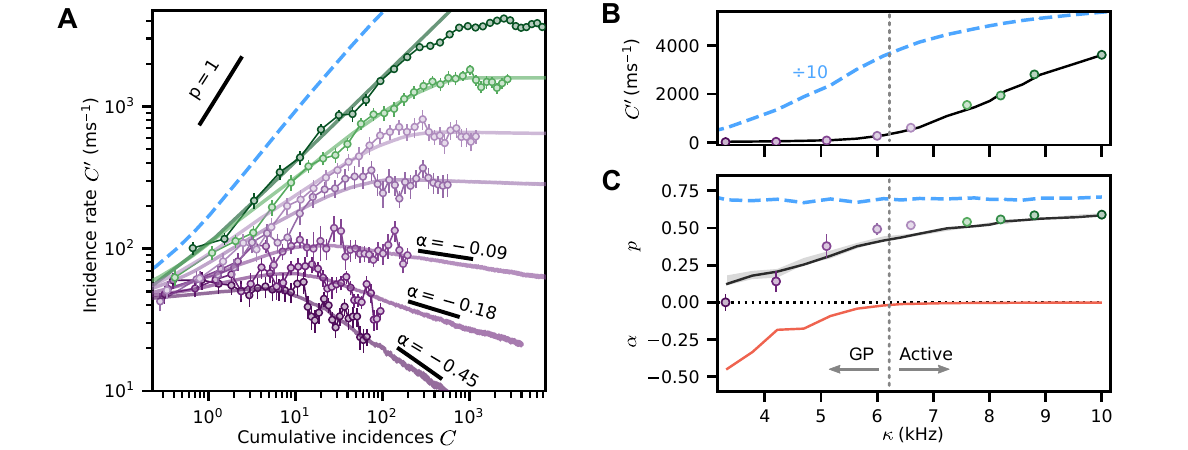}
	\caption{ \footnotesize\textbf{Rydberg excitation incidence curves for different facilitation rates showing power-law growth and Griffiths effects.} 
	\textbf{A} Incidence rate $C'$ versus cumulative incidences $C$ for different facilitation rates $\kappa =\{3.3, 4.2, 5.1, 6.0, 6.6, 7.6, 10\}\,$kHz (from purple to green). Error bars show the standard error of the mean over typically $16$ repetitions of the experiment. The straight dark-green line is a power-law fit to a representative dataset yielding the exponent $p=0.59(1)$. The solid curves are from the simulations of the SIS model on a heterogeneous network and the blue dashed line is a corresponding simulation for a locally homogeneous network for $\kappa=10\,$kHz and a comparable system size which gives $p\approx 0.67$.
    \textbf{B} Transition from a sub-critical state ($C'\approx 0$) to an active state ($C'>0$) at late times $t=2\,\mathrm{ms}$ as a function of $\kappa$. The solid-black curve and the dashed-blue curve (scaled by a factor of 10 for visual comparison) show simulations of the heterogeneous and locally homogeneous network models respectively, which exhibit different thresholds and incidence rates. \textbf{C} Characterization of the deceleration of growth parameter $p$ (experimental data points and simulations as a black line) and relaxation exponent $\alpha$ (orange line, numerical simulations only) versus $\kappa$. The vertical dashed line indicates the cross-over point between the Griffiths phase (GP) and  the active phase. Uncertainties computed from the standard deviation over $100$ bootstrap resamplings are shown as error bars except where they are smaller than the data points.
	}
	\label{fig:Cprime_C}
\end{figure*}

\subsection*{Observation of epidemic growth}
To exemplify the analogy to epidemics, in Figure~\ref{fig:atomsnetwork}D we present data for $\kappa=10\,$kHz showing different stages of the dynamics. Immediately following the seed excitation pulse we observe a period of very fast growth of the Rydberg excitation number, i.e. within the Rydberg state lifetime the excitation number increases from its initial value to more than $400$, corresponding to more than $5$ doublings in $0.19\,$ms. At around $t\approx 0.5\,$ms, after the initial growth stage, the system saturates with a high constant excitation number (i.e. an endemic state). However the saturation value is still significantly lower than the estimated maximum number of excitations that can fit in the system $\gtrsim 2000$ assuming an inter-Rydberg spacing of $\sim R_\mathrm{fac}$. On even longer timescales than those studied here ($\gtrsim 10$\,ms), the system should eventually relax back to an absorbing or self-organised critical state due to the gradual depletion of particles~\cite{HelmrichSOC,Ding2020}.

The growth phase of many real epidemics is observed to follow a characteristic power-law dependence described by the phenomenological generalized-growth model (GGM)~\cite{Viboud2016},
\begin{equation}\label{eq.1}
   C'(t) = rC^{p}(t).
\end{equation}
This describes a relation between incidence rate $C'$ and cumulative number of infections $C = \int_0^tC'(t')dt'$, where 
$r$ is the growth rate at early times and $p$ is the `deceleration of growth' which is an important parameter in classifying epidemics~\cite{Viboud2016}. Exponential growth in time is characterised by $p=1$, while $p<1$ corresponds to power-law growth $\propto t^\eta$ with $\eta=p/(1-p)$.

In Figure~\ref{fig:Cprime_C}A we represent the data from Fig.~\ref{fig:atomsnetwork}D in terms of $C'$ [instantaneous number of excitations divided by their lifetime $\tau=(2\pi\Gamma)^{-1}$] against its time integral $C$, shown by the darkest green data points. This clearly shows that the incidence rate follows the GGM over several decades (evidenced by a straight line on a double logarithmic scale) with a deceleration of growth parameter $p=0.59(1)$ that is comparable to empirical observations of real epidemics~\cite{Viboud2016}. In fact power-law growth with varying exponents $p<0.6$ is a general feature of the system dynamics, as seen in Fig.~\ref{fig:Cprime_C}A for different $\kappa$ values from $3.3\,$kHz to $10\,$kHz (depicted with different colors), together with the corresponding $p$ values plotted in Fig.~\ref{fig:Cprime_C}C as determined from fits to the initial growth stage. This is to be contrasted with exponential growth ($p=1$, solid black line with a steeper slope). We also point out that each curve saturates at a different $\kappa$-dependent value, with some curves showing evidence for slow relaxation back towards zero incidences (the lowest three curves in Fig.~\ref{fig:Cprime_C}A). In the study of epidemics, power-law growth with $p<1$ is commonly associated to a few underlying mechanisms, most prominently spatial constraints and heterogeneity in the underlying network structure~\cite{Viboud2016}. In the following we use this insight to develop a spatial network model which quantitatively describes the experimental observations and can be directly linked to the microscopic details of the system, something that is rarely possible for real epidemics.

\subsection*{Emergent heterogeneous network}
To explain the experimental observations we develop a physically motivated susceptible-infected-susceptible (SIS) network model. We assume the two-dimensional gas can be subdivided into cells that represent nodes of a network~(Fig.~\ref{fig:atomsnetwork}C). Each cell can either be in a susceptible state (absence of Rydberg excitation, $I_i=0$) or infected (one Rydberg excitation, $I_i=1$), and contains a certain number of particles $N_i$ that can be excited. Vacant cells with $N_i=0$ translate to missing nodes. The probability for a given node $i$ to become infected is described by the following stochastic master equation~\cite{PastorSatorras2015}
\begin{equation}
\frac{dE[I_i(t)]}{dt} = E\left[-\Gamma I_i(t)+\kappa N_i \biggl(1-I_i(t)\biggr)\sum_ja_{ij}I_j(t)\right],\\
\label{eq:SIS}
\end{equation}
\noindent where $E[\cdot]$ denotes the expectation value. The node weights $N_i$ and the adjacency matrix $a_{ij}$ together determine the probability for transmission of an infection from cell $j$ to $i$. In the special case $N_i=\mathrm{const.}, a_{ij}=1$, this reduces to the well studied homogeneous compartmental model~\cite{PastorSatorras2015} that exhibits exponential growth. However, spatially structured adjacency matrices can give rise to more complex spatio-temporal evolution~\cite{Watts1998collective}. 

To define the adjacency matrix $a_{ij}$ we coarse grain our system into hexagonal cells (each with area $\sim R_\mathrm{fac}^2$), corresponding to a triangular network of nodes with $a_{ij}=1$ for each of the six nearest neighbours to each node. This is motivated by the fact that hexagonal packing provides the densest possible tiling of strongly interacting Rydberg excitations in two-dimensional space~\cite{Schauss2015Crystal}, although the underlying atomic gas has no such apparent structure. 
The $N_i$ are sampled from a Poissonian distribution with a spatially dependent mean $\mu_i= \epsilon(\kappa) n_\mathrm{2d}(x_i,y_i)R_\mathrm{fac}^2$, which is set by the accessible phase space fraction for facilitated excitation $\epsilon(\kappa)<1$ (a free parameter, elaborated on below) and the value of the two-dimensional Gaussian density distribution of atoms in the trap. Thus Eq.~(\ref{eq:SIS}) describes a heterogeneous network where each node has a (spatially) fluctuating weighted degree $s_i=\sum_j a_{ij}N_j$ with a mean and variance approximately equal to $6\mu_i$. 

To numerically simulate this model we solve Eq.~(\ref{eq:SIS}) using a Monte-Carlo approach~\cite{Chotia_2008}. In each time step we compute the transition rate for each node $R_i=\nobreak\kappa N_i(1-\nobreak I_i)\sum_j a_{ij}I_j+\Gamma I_i$. One node $m$ is then picked at random according to the weights $R_i$ and its state is flipped $I_m\rightarrow 1-I_m$. The timestep is computed according to $dt = -\ln(X) / (2\pi\sum_i R_i)$, where $\ln$ is the natural logarithm and $X$ is sampled from a uniform distribution on $[0,1)$. For the initial state we consider a fixed number of $\sum_i I_i(t=0)=8$ seed excitations randomly distributed among the nodes with weights $N_i$.

The numerical simulations, shown as solid curves in Fig.~\ref{fig:atomsnetwork}D and Fig.~\ref{fig:Cprime_C}, are in excellent agreement with the experimental observations. 
Importantly they fully reproduce the fast power-law growth with $p< 0.6$, the different plateau heights and even the late-time relaxation as a function of $\kappa$. The only free parameter in the model is $\epsilon(\kappa)$ which is adjusted for each curve and is found to be a monotonically increasing function of $\kappa$ with $0.02< \epsilon(\kappa)< 0.1$ over the explored parameter range. This parameter directly controls the network structure, i.e. for $\kappa=10\,$kHz the network consists of $M\approx 2300$ nodes with $N_i>0$ and the local $s_i$ follow approximately Poissonian distributions with $\langle s_i\rangle=\mathrm{var}(s_i)\leq 5.3$ (maximal at trap center) while for $\kappa=\nobreak3.3\,$kHz, $M\approx 660$ and $\langle s_i\rangle=\mathrm{var}(s_i)\leq 1.3$. For comparison, the dashed-blue lines in Fig.~\ref{fig:Cprime_C} show comparable simulations with $N_i=\mu_i$, i.e. corresponding to a locally homogeneous network with the same average node degree. These homogeneous network simulations show faster initial growth, constant $p$ values $\approx 0.7$, higher plateaus saturating at the system size limit and a dramatic shift of the critical point to lower $\kappa$ values, which are inconsistent with the experimental data. The good agreement between experiment and heterogeneous network simulations demonstrates that the emergent macroscopic dynamics of the system crucially depend on the weighted node degree distributions and heterogeneity controlled by the atomic density and the parameter $\epsilon(\kappa)$.

The heterogeneous spatial network model described by Eq.~(\ref{eq:SIS}) provides an accurate and computationally-efficient coarse grained description of the physical system and its dynamics involving just a few microscopically controlled parameters. The importance of heterogeneity is particularly surprising since atomic motion could be expected to quickly wash out the effects of spatial disorder (the characteristic thermal velocity corresponding to the gas temperature at $20\,\mu$K is $v_{th}=65\,\mu$m/ms $\approx 3.5R_\mathrm{fac}/\tau$. Our findings can be explained by assuming that the facilitation constraint depends on both the relative positions and velocities of the atoms. Taking into account the Landau-Zener transition probability for moving atoms confirms that only atom pairs with small relative velocities $v_\mathrm{LZ}\lesssim 1\,\mu$m/ms contribute to the spreading of facilitated excitations~\cite{Helmrich2018}. This provides a qualitative explanation for the inferred $\epsilon(\kappa)\ll 1$ and its approximate $\kappa$ dependence (due to the intensity dependence of $v_\mathrm{LZ}$) [Supplementary Material]. It also sets the timescale for diffusion in phase space longer than the duration of our observations $\gtrsim 2\,$ms. Thus spatial constraints and (effectively static) heterogeneity can be understood as properties of an emergent network structure that is dynamically formed while the laser coupling is on [see also \cite{Bettelli2013} for a related interpretation of excitation dynamics on smaller preformed emergent lattices].

Spatial disorder is known to play a very important role in condensed matter systems, giving rise to new many-body phases, localization effects and glassy behavior\cite{Vojta2019}. There is still much to be explored concerning analogous effects of disorder and heterogeneity on non-equilibrium processes on networks. One key theoretical finding however is the emergence of an exotic Griffiths phase~\cite{Moreira1996,Voltja2005,Munoz2010}, expected to replace the singular critical point between the sub-critical and active phase by an extended critical phase, leading to slow relaxation and strongly modified non-equilibrium critical properties (e.g. power-law dynamical behavior with continuously varying exponents). This provides a natural explanation for several of our experimental observations. First of all, the relatively short time for each curve to reach the plateau and the strong $\kappa$ dependence of the plateau heights are compatible with the presence of rare regions with an over average infection rate that span only a fraction of the entire system, controlled by the disorder strength entering via $\epsilon(\kappa)$. This also explains the sizable shift of the critical point between sub-critical ($C^\prime \approx0$) and active ($C^\prime >0$) phases to higher values of $\kappa$ as compared to the expectation for a locally homogeneous system seen in Fig.~\ref{fig:Cprime_C}B. Finally, we point out the slow relaxation of the sub-critical curves in Fig.~\ref{fig:Cprime_C}A. These curves are compatible with power-law decays with disorder dependent relaxation exponents $\alpha<0$, which is the defining characteristic of the Griffiths phase~\cite{Munoz2010}. While these experiments were limited to relatively short times $<2\,$ms (to minmize the impact of particle loss), the numerical simulations confirm power-law relaxation over two orders of magnitude in time, depicted by solid lines in Fig.~\ref{fig:Cprime_C}A, with corresponding $\alpha\leq 0$ exponents shown in Fig.~\ref{fig:Cprime_C}C. On this basis we find that power-law growth (with $0.5\leq p\leq 0.6$) is associated with the transition from a Griffiths phase to an active phase (for $\kappa>6\,$kHz coinciding with $\alpha\approx 0$), whereas the disorder-free absorbing state phase transition is expected for $\kappa\lesssim \Gamma$~\cite{Munoz2010}.

\subsection*{Conclusion}
This work highlights a controllable physical platform for experimental network science situated at the interface between simplified numerical models and empirical observations of real-world complex dynamical phenomena. Ultracold atoms provide the means to introduce and control different types of reaction-diffusion processes as studied here, but also to realize different types of spatial networks by structuring the trapping fields~\cite{WangTweezers} and to access the full spatio-temporal evolution of the system~\cite{Schauss2015Crystal}. Our discovery that the growth dynamics of a driven-dissipative atomic gas is described by an emergent heterogeneous network that is relatively robust to particle motion suggests similar effects could also be observable in noisy room temperature environments~\cite{Urvoy2015,Ding2020}. Thus heterogeneous network dynamics and Griffiths effects may arise naturally in very different non-equilibrium systems, having important implications, for example, in understanding non-equilibrium criticality without fine tuning~\cite{Griffithsbrain,HelmrichSOC,Ding2020} and for finding effective strategies for controlling dynamics on complex networks~\cite{Buono2013}.

\subsection*{Acknowledgments}
We acknowledge valuable discussions with C\'edric Sueur. This work is supported by the `Investissements d'Avenir' programme through the Excellence Initiative of the University of Strasbourg (IdEx), the University of Strasbourg Institute for Advanced Study (USIAS) and is part of and supported by the DFG SPP 1929 GiRyd and the DFG Collaborative Research Center `SFB 1225 (ISOQUANT)'. T.M.W., S.S. and M.M. acknowledge the French National Research Agency (ANR) through the Programme d'Investissement d'Avenir under contract ANR-17-EURE-0024. M.M. acknowledges QUSTEC funding from the European Union's Horizon 2020 research and innovation programme under the Marie Sk\l{}dowska-Curie grant agreement No. 847471.  S.D. acknowledges support by the European Research Council (ERC) under the Horizon 2020 research and innovation program, Grant Agreement No. 647434 (DOQS).



\section*{Supplementary material}

# Supplementary Materials for
# Epidemic growth and Griffiths effects on an emergent network of excited atoms

T. M. Wintermantel[1,2], M. Buchhold[3], S. Shevate[2], M. Morgado[2],
Y. Wang[2], G. Lochead[2], S. Diehl[3], S. Whitlock[2*]

[1]Physikalisches Institut, Universität Heidelberg, 69120 Heidelberg, Germany
[2]ISIS (UMR 7006), University of Strasbourg and CNRS, 67000 Strasbourg, France
[3]Institut für Theoretische Physik, Universität zu Köln, 50923 Cologne, Germany

*Corresponding author. E-mail: whitlock@unistra.fr.

July 15, 2020

## 1 Materials and Methods

### 1.1 Experimental sequence and calibration of parameters

The experimental procedure consists of three main steps: (i) Initially a small number of seed excitations are prepared at random positions in the gas. For this we keep the laser frequency fixed at $\Delta = -30$ MHz below the zero field resonance and briefly applying an electric field of $0.28$ V/cm for $4\,\mu$s, exploiting the DC Stark effect to tune the atoms into resonance. The laser is then momentarily switched off for $6\,\mu$s to ensure the electric field is fully off before starting the off-resonant driving. (ii) Next we apply the off-resonant laser field which causes rapid growth of the number of excitations in the gas. We calibrate the single-atom facilitation rate $\kappa$ against a measurement of the initial growth rate $r = 27(8)$ kHz, measured for high intensity and very short times $t \ll \tau$ where many-body effects can be safely neglected. This is then divided by an

1

estimate of the (cloud averaged) mean number of particles that meet the facilitation condition $\bar{\mu} = 2.7$ assuming each seed excitation is isolated. The latter is estimated from the detailed experiment-theory comparison to the spatial SIS model presented in the manuscript. (iii) After a variable exposure time $t$ we measure the total number of excitations in the gas. For this we switch on a large electric field to ionize the Rydberg states and guide the ions onto a microchannel plate (MCP) detector. The conversion factor from MCP voltage to the number of Rydberg excitations is calibrated against an independent absorption measurement of the number of particles removed from the gas after a long exposure assuming each Rydberg excited atom is eventually lost from the trap with rate $\Gamma$.

## 1.2 Landau-Zener probability for facilitated excitation

By comparing the SIS network simulations to the data we infer that the fraction of atoms that participate in the excitation dynamics is relatively small. This is quantified by the fitted $\epsilon(\kappa)$ values that vary between 0.023 and 0.094 (for $\kappa = 3.3\,\text{kHz}$ and $\kappa = 10\,\text{kHz}$ respectively). These small values of $\epsilon$ and the approximate $\kappa$ dependence can be explained by the velocity dependence of the Landau-Zener transition probability, which restricts facilitation to atoms with small relative velocities $v \lesssim v_{\text{LZ}} \ll v_{\text{th}}$ [for a related calculations see Appendix E in (*1*)]. The Landau-Zener velocity can be expressed as $v_{\text{LZ}} = \pi^2 \Omega^2 / \dot{V}$, where $\Omega$ is the light-matter coupling strength and $\dot{V}$ is the slope of the Rydberg-Rydberg interaction potential evaluated at the facilitation radius (*1*). $\epsilon$ can be understood as the number of atoms that can be facilitated in a neighbouring cell within the Rydberg state lifetime divided by the mean number of atoms in each cell $n_{2d} R_{fac}^2$. The flux of atoms passing through a 1/6 segment of the facilitation shell is $\Phi = \pi R_{fac} n_{2d} v_{th}/3$. However, only a fraction of these atoms $f_v \approx v_{\text{LZ}}/\sqrt{\pi} v_{\text{th}}$ fulfill the Landau-Zener condition with relative velocity $|v| < v_{\text{LZ}}$. Combining the above gives $\epsilon = \Phi \tau f_v / n_{2d} R_{fac}^2 \approx \sqrt{\pi} \tau v_{\text{LZ}}/(3 R_{fac})$.

For realistic experimental parameters $R_{\text{fac}} = 3.5\,\mu\text{m}$, $\dot{V} = 1 \times 10^5\,\text{kHz}\,\mu\text{m}^{-1}$ (*2*) and $\Omega \sim 100\,\text{kHz}$, we find $v_{\text{LZ}} = 1\,\mu\text{m/ms}$. This is small compared to the thermal velocity $v_{\text{th}} = 65\,\mu\text{m/ms}$ for $T = 20\,\mu\text{K}$. Inserting these parameters into the expression above for the phase space fraction yield $\epsilon = 0.03 \left(\frac{\Omega}{100\,\text{kHz}}\right)^2$. This simple estimate falls within the range of values inferred from the experiment-theory comparison, even though it still does not account

for all the microscopic experimental details, e.g., multiple excitation resonances associated to Zeeman substructure or possible mechanical forces between the atoms.

# 2 Supplementary Text

## 2.1 Extracting the growth and relaxation parameters

Fig. S1A shows the complete experimental data-set up to $2\,\mathrm{ms}$ off-resonant laser excitation and for kappa values of $\kappa = \{3.3, 4.2, 5.1, 6.0, 6.6, 7.6, 8.2, 8.8, 10\}$ kHz, color-coded from purple to green. The solid lines are the best fit numerical simulation results of the susceptible-infected-susceptible (SIS) network model as described in the manuscript using $\epsilon(\kappa) = \{0.023, 0.027,$ $0.037, 0.056, 0.067, 0.078, 0.083, 0.088, 0.094\}$.

To extract the growth and relaxation parameters from this data, we extend the generalized growth model (GGM) to allow for different exponents in the growth and relaxation phases

$$\mathcal{C}'(t) = r\mathcal{C}^p(t) \left[1 + \left(\frac{C(t)}{K}\right)^{\frac{p-\alpha}{\beta}}\right]^{-\beta}. \tag{1}$$

In this equation $p$ is the deceleration of growth parameter, $\alpha$ is the exponent for the late time recovery phase. $K$ and $\beta$ determine the location and sharpness of the crossover. Curves with $\alpha < 0$ will eventually recover (i.e. number of excitations decreases to zero) while $\alpha = 0$ describes an endemic state. The endemic state is characterised by a constant number of excitations $C' = rK^p$. The number of excitations at the crossover point $(C = K)$ is $C' = rK^p/2^\beta$.

Figure S1B shows the incidence rate $C'$ against its time integral $C$ for the full experimental data set and fits of the extended GGM of eq. 1 (solid lines). We observe good agreement between the data and the fit curves. The $p$ parameters corresponding to these fit results are presented in Fig. 2C in the manuscript.

## 2.2 Characterization of the emergent network structure

Finally we present a statistical analysis of the emergent network structure inferred from the SIS simulations that reproduce the experimental data. In Fig. S2A,B we present the weighted degree

3

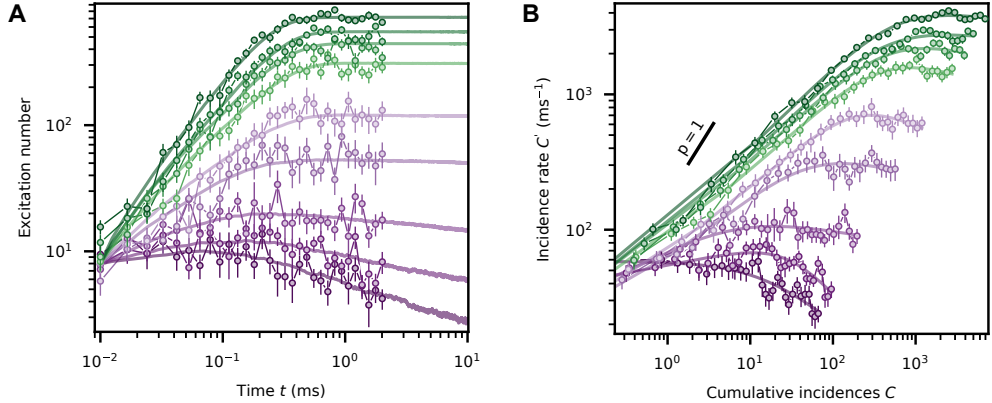

Figure S1: **Extracting the growth and relaxation parameters on the full data set. A** Number of instantaneous excitations as a function of excitation time $t$ showing all the raw experimental data (disks), where the colors from purple to green correspond to increasing facilitation rates $\kappa$ values of $\{3.3, 4.2, 5.1, 6.0, 6.6, 7.6, 8.2, 8.8, 10\}$ kHz. The solid lines result from simulations using the described network SIS model. **B** Incidence rate $C'$ versus cumulative incidences $C$. The data points (disks) are obtained from the data shown in A. The solid lines show the fit results of the extended generalized growth model (see text for more details). The error bars of the measurement data (disks) represent the standard error of the mean over typically 16 experimental runs.

distributions, averaged over the full network and over 1000 realizations, for a sub-critical and super-critical network for $\epsilon(\kappa = 3.3\,\mathrm{kHz}) = 0.023$ and $\epsilon(\kappa = 10\,\mathrm{kHz}) = 0.094$ respectively. As for the simulations presented in the main paper, the system size is limited to a maximum of 10000 nodes. The weighted degree distributions are qualitatively similar, following an approximately Gaussian dependence as evidenced by parabolic shape on a semilog scale. However these networks are very different, as evidenced by the connected component or cluster size distribution. A cluster is defined as a set of nodes where each node one can reach any other node via edges, and the cluster size is the number of nodes in this set. Starting from a single seed excitation, the maximum number of excitations will be constrained by the size of its cluster. Fig. S2C shows that the cluster size distribution for the sub-critical network with $\kappa = 3.3\,\mathrm{kHz}$ follows an approximately exponential dependence, i.e. large connected clusters are exponentially rare. In contrast, the supercritical network with $\kappa = 10\,\mathrm{kHz}$ exhibits much larger connected clusters with a peak in the cluster size distribution around 750 (Fig. S2D).

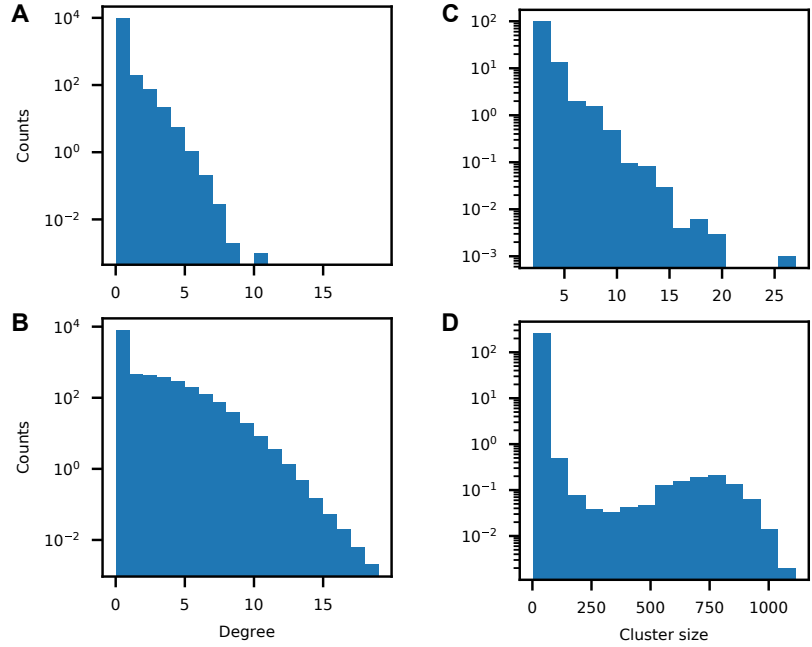

Figure S2: **Characterization of the underlying structure of the network used in the SIS model simulations.** Weighted degree distributions for **A** $\kappa = 3.3\,\text{kHz}$ (sub-critical) and **B** $\kappa = 10\,\text{kHz}$ (super-critical). Cluster size distribution for **C** $\kappa = 3.3\,\text{kHz}$ and **D** $\kappa = 10\,\text{kHz}$.



\end{document}